\documentclass[11pt,twoside]{article}
\usepackage{asp2010}

\resetcounters

\begin{document}

\title{The many faces of Betelgeuse}

\author{Vikram~Ravi$^{1}$, Ed~Wishnow$^{1}$, Sean~Lockwood$^{1}$, Charles~Townes$^{1}$}
\affil{$^{1}$Space Sciences Laboratory and Department of Physics, University of California, Berkeley, CA 94720, USA}

\begin{abstract}

The dynamics of the surface and inner atmosphere of the red supergiant star Betelgeuse are the subject of 
numerous high angular resolution and spectroscopic studies. Here, we present three-telescope 
interferometric data obtained at 11.15\,$\mu$m wavelength with the Berkeley Infrared Spatial Interferometer (ISI), 
that probe the stellar surface continuum. We find striking variability in the size, effective temperature, 
and degree of asymmetry of the star over the years 2006$-$2009. These results may indicate an 
evolving shell of optically thick material close to the stellar photosphere.

\end{abstract}

\section{Introduction}

The red supergiant star Betelgeuse ($\alpha$ Orionis) is a key target for high 
angular resolution imaging studies at optical and infrared wavelengths because of 
its large angular diameter \citep[$\sim$0.047\arcsec,][]{mp21} and high apparent magnitude 
($M_{K}=-4.38$). In recent years, a variety of interferometric campaigns have revealed 
up to three largely unresolved spots on the stellar surface in various visible, near-infrared and 
mid-infrared bands \citep[e.g.][]{wdh97,hpl+09,tcw+07}. These spots range in intensity from a 
few to a few tens of percent of the total stellar flux. The spots appear to be transient, with stability 
timescales of less than 8 weeks \citep{wdh97}. The properties of these spots have led to a consensus that 
they indicate large-scale convection cells, as predicted by \citet{s75}.

Numerous puzzles remain regarding the effective temperatures of the spots, and 
their possible roles in mass loss from Betelgeuse and in shaping the asymmetric inner atmosphere 
of the star. Imaging studies in the ultraviolet \citep{gd96} suggest the presence of a plume 
of material at chromospheric temperatures. Wide-band radio images 
of Betelgeuse on spatial resolutions of a few stellar radii ($R_{*}$) also show asymmetry in 
largely neutral gas at temperatures of a few thousand Kelvin \citep{lcw+98}. A plume at 6$R_{*}$ 
was also observed by \citet{kvr+09}, possibly associated with the same gaseous component observed by 
\citet{lcw+98}. Near- and mid-IR spectro-interferometric data for Betelgeuse have been 
modeled by a variety of authors using a dense shell of molecular gas at temperatures of 
$\sim$2000\,K located above a 3600\,K photosphere \citep{prc+04}, mainly composed of H$_{2}$O 
\citep{t00,pvr+07,ohb+09}. This shell, known as a `MOLsphere', is generally thought to lie 
between 1.3$R_{*}$ and 1.5$R_{*}$, and has been shown to be patchy \citep{ohb+09}. Evidence for MOLspheres is 
also seen in other stars \citep[e.g. T Lep,][]{lmm+09}. Betelgeuse is further known to possess an extended 
dust envelope, with two shells at angular radii of approximately 1\arcsec and 2\arcsec~\citep{dbd+94}. A third, 
newly formed component was found by \citet{bdh+96} at 0.1\arcsec. The mechanisms that drive the formation of the 
dust and shape the dynamic regions close to the surface of Betelgeuse are not well known.

\section{Observational methods and results}

The ISI is a three-telescope interferometer with heterodyne detection systems at each telescope 
\citep[Wishnow et al., these proceedings;][]{wmr+10}. 
This enables measurements over a narrow bandwidth of $\sim$5.4\,GHz 
centered on a given CO$_{2}$ laser local oscillator wavelength. The 11.15\,$\mu$m wavelength chosen for these 
measurements corresponds to a region that is free of prominent stellar molecular lines \citep{wht03}. 

Visibilities and closure phases at spatial frequencies between 
20$-$37\,SFU\footnote{1 Spatial Frequency Unit (SFU) = $1\times10^{5}$ cycles/radian.} 
were obtained over timespans of a few weeks in each year between 2006 and 2009 
(see Table 1 for an observing log). The van Cittert-Zernicke theorem relates the complex cross-correlation 
between the detected electric fields at different telescopes $-$ the complex visibility $-$ 
to the two-dimensional Fourier transform of the source brightness distribution. The ISI, like all optical/IR 
interferometers, however measures the Michelson fringe visibility, which is the power at a given spatial 
frequency normalized by the total source power. The closure phase is the sum of the fringe phases measured 
in each of the three baselines. While individual visibility 
phases vary due to atmospheric turbulence, the closure phase is independent of these 
variations \citep{j58}. The spatial frequency sampled by a telescope baseline at a given time is proportional to 
$b/\lambda$, where $b$ is the baseline length orthogonal to the source direction and $\lambda$ is the 
wavelength. By obtaining measurements approximately every five minutes, the ISI samples visibilities 
and closure phases over a range of spatial frequencies and position angles as the Earth rotates. 
The telescope configuration used for the measurements reported here was an equilateral triangle 
with $\sim$35\,m baseline lengths.

\begin{table}[h]
\begin{center}
\caption{Observing log.}
\smallskip
{\small
\begin{tabular}{cc}
\tableline
\noalign{\smallskip}
Year & Dates (UTC) \\
\noalign{\smallskip}
\tableline
\noalign{\smallskip}
2006 & 8, 9, 10 Nov; 7 Dec \\
2007 & 14, 15, 16 Nov \\
2008 & 22, 23, 24 Sep \\
2009 & 7 Oct; 4, 5, 7, 9, 18, 20, 22 Nov \\
\noalign{\smallskip}
\tableline
\end{tabular}
}
\end{center}
\end{table}

\begin{figure}[!ht]
\plotone{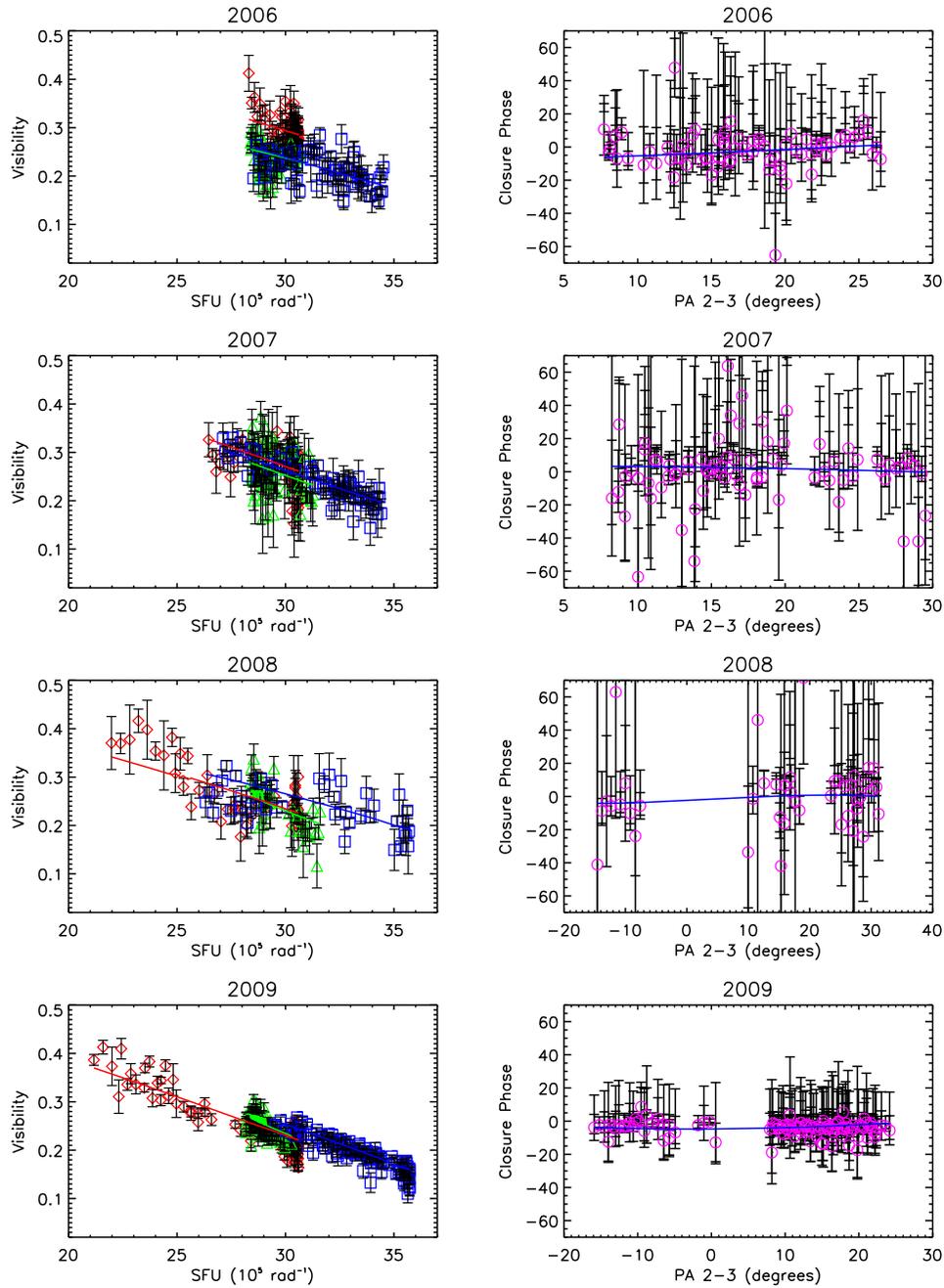}
\caption{Visibilities (left column) and closure phases (right column) from the 2006, 2007, 2008 and 2009 epochs. 
Data in red, green and blue (diamonds, triangles and squares respectively) denote telescope baselines 
1-2, 2-3 and 3-1, respectively. The closure phases are plotted against the position angle of the 2-3 
baseline East of North. The solid lines indicate the best-fit image models.}
\end{figure}

The visibility and closure phase data for each epoch are shown in Figure 1. The data were calibrated using 
nightly observations of Aldebaran ($\alpha$ Tauri), which was assumed to have a diameter of 20\,mas 
and zero closure phase. Data that were affected by systematic errors and poor atmospheric conditions were 
discarded. The data in each epoch were fit using a simple image model of a centered uniform stellar disk, a 
point source at an arbitrary location and a large, spherically-symmetric dust shell that did not 
contribute to the visibility at the sampled spatial frequencies. The fits used a weighted least-squares 
minimization technique, with the weights calculated as the inverse squares of the errors. Each baseline and the 
closure phase data were weighted equally. For a baseline sampling the two-dimensional 
spatial frequency plane at a point ($u$, $v$), the model complex visibility was
\begin{equation}
V(u,v)=\frac{2AJ_{1}(2\pi r\sqrt{u^{2}+v^{2}})}{2\pi r\sqrt{u^{2}+v^{2}}}+Pe^{-2\pi i(ux+vy)},
\end{equation}
where $A$ is the fraction of the total flux contributed by the stellar surface, $J_{1}$ denotes 
a Bessel function of order unity, $r$ is the stellar angular radius, $P$ is the fraction of the total flux 
contributed by the point, and $x$ and $y$ are the angular coordinates of the point with respect to the 
uniform disk center.
The fit results are presented in Table 2, with the errors in the last decimal places given in parantheses. The 
total flux density, $F_{\nu}$, of the star and dust was also estimated for each epoch assuming a flux 
density of 615\,Jy for Aldebaran \citep{mgd98}.

\begin{table}[!ht]
\begin{center}
\caption{Fit results.}
\smallskip
{\small
\begin{tabular}{ccccccc}
\tableline
\noalign{\smallskip}
Year & $A$ & $r$ (mas) & $P$ & $x$ (mas) & $y$ (mas) & $F_{\nu}$ ($10^{3}$ Jy) \\
\noalign{\smallskip}
\tableline
\noalign{\smallskip}
2006 & 0.51(2) & 24.5(8) & 0.05(1) & -2.42(1) & -23.70(4) & 4.2(2)\\
2007 & 0.55(2) & 25.7(9) & 0.02(1) & -3.25(2) & 16.80(8) & 4.1(3) \\
2008 & 0.47(2) & 24.6(8) & 0.04(2) & -14.30(3) & 12.75(3) & 3.8(5) \\
2009 & 0.540(5) & 25.9(1) & 0.013(1) & -24.40(1) & 13.92(1) & 4.4(1) \\
\noalign{\smallskip}
\tableline
\end{tabular}
}
\end{center}
\end{table}

Figure 2 shows the best-fit model images for each epoch. The 2006 data were previously analyzed by 
\citet{tcw+07}, and despite slightly different techniques used here, the results closely match. 
We find point sources of varying intensity in each epoch, as well as significant 
changes in the uniform disk radius of the star. Limb darkening is estimated to reduce the apparent 
stellar diameter by less than 1$\%$ at our wavelength \citep{bdh+96}, and hence was not accounted for. 

\begin{figure}[!ht]
\plotone{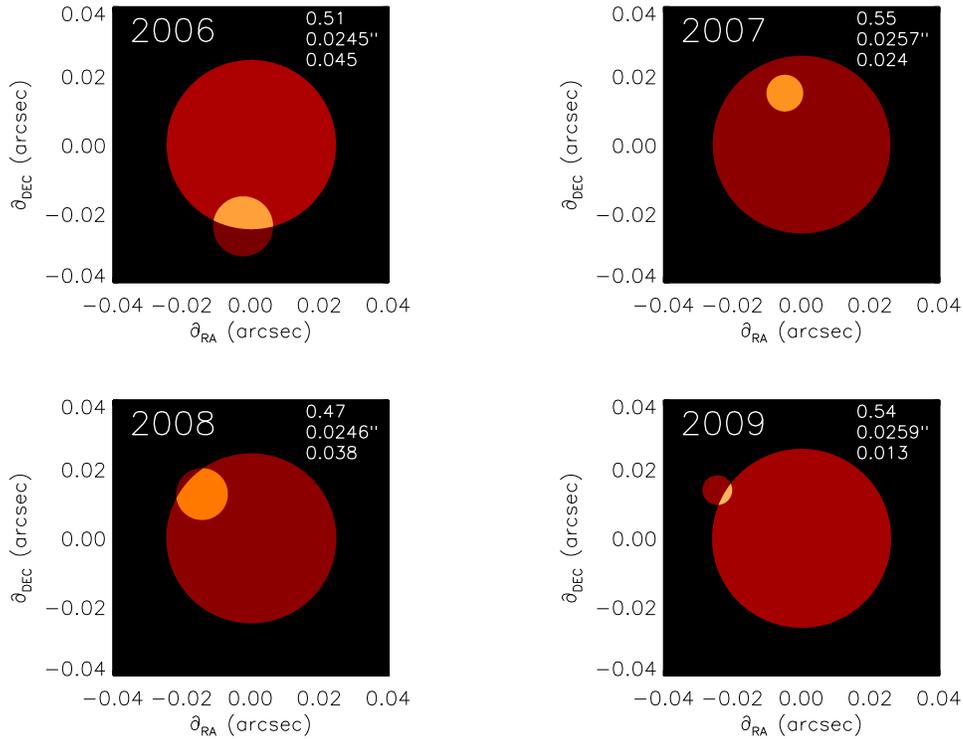}
\caption{Depictions of the best-fit image models for Betelgeuse during each of the 2006, 2007, 2008 and 
2009 epochs. Each figure includes the fit parameters: the fraction of the total flux from the star, the stellar 
radius in arcseconds, and the fraction of the total flux from the point. The point sources have been given the 
uniform disk sizes that they would have if they represented regions at a temperature of 7200\,K. Our 
upper limit on the point source diameter is 20\,mas.}
\end{figure}

\subsection{The effective temperature of the stellar surface}

Our absolute measurements of the total flux of Betelgeuse, combined with the fitted stellar size and fraction 
of the total flux, enable an estimate of the effective temperature of the stellar surface. 
We used a Planck function, integrated over a 5.4\,GHz bandwidth centered on 11.15\,$\mu$m, to 
estimate the temperature of the stellar disk. These results are presented in Figure 3. 
We did not account for extinction caused by the interstellar medium or by intervening 
layers of dust. However, the interstellar medium and the dust together are expected to attenuate the 
true stellar flux by only $\sim$3.1$\%$ \citep{dbd+94}. We also did not account for emission from 
the dust along the line of sight to the stellar face, again because of the low optical depth. 

\begin{figure}[!ht]
\plotone{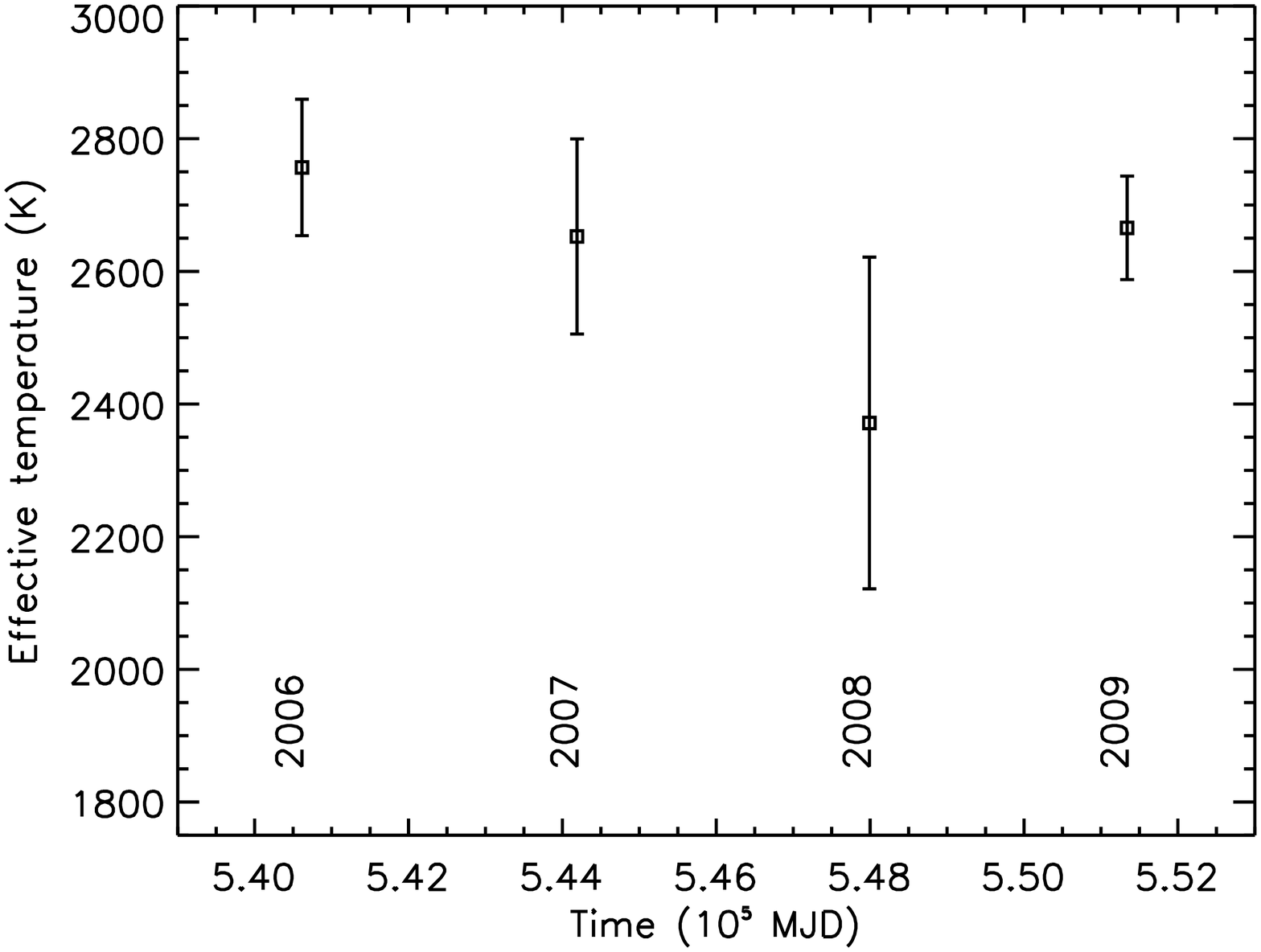}
\caption{Estimated effective temperatures of the stellar surface of Betelgeuse, neglecting the contribution 
of any asymmetry, measured over the 2006, 2007, 2008 and 2009 epochs.}
\end{figure}

\section{Discussion}

\subsection{The source of continuum opacity in the mid-IR}

In order to correctly interpret our results, it is crucial to elucidate the nature of the continuum 
opacity probed in our 11.15\,$\mu$m observations. This is not an easy task. 
The `surface' of a star at a given wavelength can be considered to be the location where the optical depth 
is unity. The stellar photosphere can be thought of as a surface where the opacity is equivalent to 
the Rosseland mean opacity, which is a frequency average of the monochromatic opacity. 
\citet{wht03b} concluded that the mid-IR  continuum opacity for solar abundance gas 
closely matches the Rosseland opacity under the stellar surface conditions present in red supergiant 
stars. This leads to the implicit conclusion that our 11.15\,$\mu$m observations directly probe the stellar 
photosphere. This conclusion is further supported by the fact that ISI observations in 
a wavelength band inclusive of a strong H$_{2}$O molecular line revealed a larger apparent stellar diameter 
for the AGB star Mira ($o$ Ceti) than that measured at 11.15\,$\mu$m. This was indicative of a warm water 
shell that appeared transparent at 11.15\,$\mu$m \citep{wht03}. \citet{pvr+07} further show that H$_{2}$O 
molecular lines do not contribute significantly to the 11.15\,$\mu$m opacity in their model for the 
Betelgeuse MOLsphere. We can tentatively conclude that the contribution of a continuum of water lines 
from a shell above the stellar surface to the 11.15\,$\mu$m opacity in Betelgeuse is not significant.

The contribution of the MOLsphere to the 11.15\,$\mu$m opacity through means other than water lines cannot 
however be neglected. \citet{vdv+06} suggested that amorphous alumina (Al$_{2}$O$_{3}$) dust in the process 
of formation in the MOLsphere might contribute to the opacity. Furthermore, \citet{pvr+07} find, through 
modeling of spectro-interferometric data  in the mid-IR obtained using the VLTI/MIDI instrument, that alumina as 
well as gaseous SiO appear to contribute to the opacity of the MOLsphere. At 11.15\,$\mu$m wavelength, this 
model suggests an optical depth of $\sim$2, with the opacity dominated by alumina.

An extension of the ideas of \citet{rm97} for the radio opacity of red supergiant and AGB stars 
by \citet{tt06} shows that electron-hydrogen collisions in thermally ionized 
inner atmospheres could contribute significantly to the mid-IR opacity. This opacity is 
proportional to the temperature, naively implying that if a star became hotter, its apparent size 
would increase. Our present results provide evidence in support of such a phenomenon: the smallest size 
measurement corresponds to the lowest effective surface temperature (see Table 2 and Figure 3), and the 
larger size measurements correspond to higher effective surface temperatures.

\subsection{The origin of the spots}

A clear link has been made between spots observed at optical/near-IR wavelengths and 
the giant convection cell hypothesis of of \citet{s75}, as modeled by 
\citet{chy+10}. However, it is harder to consider the spots we observe in the 
mid-IR as directly caused by high-temperature regions on the stellar photosphere related 
to convection cells. The spots seen at 700\,nm wavelength by \citet{wdh97}, which contributed 
$\sim$20$\%$ of the total flux, were attributed to regions with temperature excesses of at least 
600\,K from a mean photospheric temperature of 3600\,K. The upper limit on the spot 
diameters was 15\,mas. In contrast, Figure 2 shows that temperatures of at least 
twice the putative photospheric temperature of 3600\,K are in some cases required to produce 
spots of similar size that are consistent with our mid-IR observations. Such temperatures are 
inconsistent with the predictions of \citet{s75} for temperature fluctuations of $\pm1000$\,K on 
red giant surfaces. 

\section{Conclusions and future work}

Our present mid-IR continuum results for Betelgeuse show variability in the apparent stellar diameter, 
location and strength of asymmetries modeled as point sources, and in the effective stellar 
surface temperature, all on timescales of a year. Intriguingly, we find low effective surface temperatures, 
$\sim$1000\,K below those usually associated with the photosphere, and an approximate scaling of the 
apparent radius with temperature. These results suggest that our measurements are dominated by the behaviour 
of cool, optically thick material above the stellar photosphere. 
A possible interpretation of our temperature and size results is that we have witnessed changes in this 
shell. This interpretation could be extended to explain the systematic decrease in the apparent 11.15\,$\mu$m 
size of Betelgeuse reported by \citet{twh+09} over the interval 1993$-$2008. The dominant 11.15\,$\mu$m 
opacity source in the shell could, as suggested by \citet{vdv+06} and \citet{pvr+07}, be alumina dust. 
If electron-hydrogen collisions, however, dominate the opacity, the apparent size changes could be attributed to 
temperature changes in largely neutral gas in thermal equilibrium at our measured effective surface temperatures.

Clearly, further work is required to match our observations to radiative transfer models involving 
layers of material above the stellar photosphere, fully including the various possible opacity sources. More detailed 
modeling of the closure phase data is also necessary. Our modeled off-center point sources cannot be spots of 
similar size to those observed at optical/near-IR wavelengths, because the required surface temperatures are 
unrealistic. The observed asymmetries in our data could possibly indicate large-scale asymmetry in the 
observed surface.

\acknowledgements

W. Fitelson, B. Walp, C.S. Ryan, D.D.S. Hale, K. Tatebe, A.A. Chandler, K. Reichl, R.L. Griffith, 
V. Toy, as well as many undergraduate researchers all participated in these observations, and their 
excellent help is greatly appreciated. This research made use of the SIMBAD database. We are grateful 
for support from the Gordon and Betty Moore Foundation, the Office of Naval Research, and the 
National Science Foundation.

\bibliography{ravi_v}

\end{document}